\documentclass[twocolumn,aps,amsmath,amssymb,showpacs,pre]{revtex4}
\usepackage{graphicx}
\usepackage{epsfig}
\bibstyle{apsrev.bib}

\begin{document}
\input epsf.sty

\title{Scaling behavior of the contact process in networks with long-range connections}

\author{R\'obert Juh\'asz}
 \email{juhasz@szfki.hu} 
\affiliation{Research Institute for Solid
State Physics and Optics, H-1525 Budapest, P.O.Box 49, Hungary}
\author{G\'eza \'Odor}
 \email{odor@mfa.kfki.hu} 
\affiliation{Research Institute for Technical Physics and Materials 
Science, H-1525 Budapest, P.O.Box 49, Hungary}

\date{\today}

\begin{abstract}
We present simulation results for the contact process on regular, cubic
networks that are composed of a one-dimensional lattice and a
set of long edges with unbounded length. 
Networks with different sets of long edges are considered, that are
characterized by different shortest-path dimensions and random-walk
dimensions.   
We provide numerical evidence that an absorbing phase transition
occurs at some finite value of the infection rate and  
the corresponding dynamical critical exponents depend 
on the underlying network.
Furthermore, the time-dependent quantities exhibit log-periodic 
oscillations in agreement with the discrete scale invariance
of the networks. 
In case of spreading from an initial active seed, 
the critical exponents are found to depend on the location of 
the initial seed and 
break the hyper-scaling law of the directed percolation universality class
due to the inhomogeneity of the networks.
However, if the cluster spreading quantities are averaged over initial
sites the hyper-scaling law is restored. 
\end{abstract}


\maketitle

\newcommand{\bc}{\begin{center}}
\newcommand{\ec}{\end{center}}
\newcommand{\be}{\begin{equation}}
\newcommand{\ee}{\end{equation}}
\newcommand{\beqn}{\begin{eqnarray}}
\newcommand{\eeqn}{\end{eqnarray}}

\section{Introduction}

It is known both for equilibrium and nonequilibrium systems 
that the presence of long-range interactions leads to different
critical behaviors compared to universality classes characteristic for
systems with short-range interactions \cite{odor}. 
This has been demonstrated for a paradigmatic model exhibiting a phase
transition to a single absorbing state, the contact process (CP), 
that has been introduced to model epidemic spreading without 
immunization \cite{ContactProcess,GrasTor}.
In this simple model, lattice sites have two states (active or
inactive) and active sites may become inactive or may render neighboring
inactive sites active. 
The absorbing phase and the active phase of the systems are 
separated by a continuous
phase transition with the universal behavior of the
directed percolation (DP) \cite{DPuni,DPuni2,DickMar}. 
Long-range interactions have been included by allowing the activation 
process for far-away inactive sites with probabilities that decay
algebraically with the distance. The critical exponents of the corresponding 
absorbing phase transition have been found to depend continuously 
on the exponent controlling the decay of activation probabilities 
\cite{hinrichsen,Hbook}.

An alternative way of realizing long-range interactions is when the dynamical
process is defined on a network with long links that connect distant sites.  
Networks of this type are the scale-free networks constructed by preferential
attachment \cite{Barab}, where the critical behavior of CP is controlled by the
degree-distribution \cite{psv,dorog}.
Other models of networks are those which are composed of a 
$d$-dimensional regular lattice and additional long edges. 
These arise e.g. in sociophysics \cite{Barab} or in the context of
conductive properties of linear polymers with crosslinks that connect remote
monomers \cite{cc}. In general, a pair of nodes separated by the 
distance $l$ is connected by an edge with a probability $p_l\simeq\beta
l^{-s}$ for large $l$ \cite{nw,moukarzel,monasson,jespersen,bb,sc,mam,coppersmith,kleinberg,bh,boettcher,juhasz}. 
We mention that the case $s=0$ corresponds to the Watts-Strogatz graph  
\cite{ws}, that displays the small-world phenomenon, although that model is
constructed by rewiring edges rather than adding new ones 
therefore the resulting graph may be disconnected.   
An intriguing property of these graphs for $d=1$ 
is that in the marginal case $s=2$,
the intrinsic properties show power-law behavior and the corresponding
exponents vary continuously with the prefactor $\beta$.
Indeed, this has been conjectured for the diameter $D$ as a
function of the number of nodes $L$, that means $D(L)\sim L^{d_{\rm min}}$ where the dimension 
$d_{\rm min}$ depends on $\beta$ \cite{bb}. Later, power-law bounds have been
established for $D(L)$ \cite{coppersmith}. 
For a class of cubic networks with $s=2$, the algebraic growth of the
diameter has  been explicitly demonstrated \cite{juhasz}. 
Moreover, the mean-square displacement of random walks in such
networks has been found to grow algebraically in time with an
anomalous random walk dimension that is characteristic for the
underlying network. This behavior contrasts with L\'evy-flights in 
the respect that, here, 
the decay exponent $s$ does not exclusively determine the diffusion exponent
but the latter depends also on the details of the structure of networks 
if $s=2$.  
As opposed to random walks, much less is known for interacting many-particle
systems on such networks \cite{boettcher}.    
In particular, the behavior of nonequilibrium systems possessing
an absorbing phase transition has not cleared up yet. 
The aim of the present work is to investigate the contact process on these 
networks. 
On the basis of the scaling of diameter and mean-square displacement
of random walks, we expect a nonequilibrium system possessing an
absorbing phase transition, such as the contact process, to be 
characterized by altered critical exponents when defined on such
networks compared to the corresponding one-dimensional model.
We will demonstrate by Monte Carlo simulations that this is indeed the
case for a class of regular networks which, concerning the diameter 
and random walks, are known to be described by power-laws, similar to
$s=2$ random networks.
The advantage of studying regular networks is that, at least, the diameter
exponent and the random-walk dimension are exactly known here and unlike for
random networks no disorder (sample) average is needed to carry out here. 

The rest of the paper is organized as follows. The networks to be
investigated are defined in section \ref{secnetworks}. The model and
the studied quantities are specified in section \ref{seccp}. 
Results of numerical simulations are presented in section
\ref{secresults} and discussed in section \ref{secdiscussion}. 

\section{Networks with long links}
\label{secnetworks}
\subsection{Definition of networks}

In this section, we specify the networks on which the contact process is
studied. 
First, a regular, one dimensional lattice (periodic or open) with $L$ sites is
considered, where the lattice sites are numbered consecutively 
from $1$ to $L$. 
To this lattice, where the degree of sites, i.e. the number
of edges emanating from a site, is $2$, links are added one by one
until all sites become of degree $3$. 
Sites of degree $2$ will be called in brief {\it free sites}, 
and $k$ will denote a fixed positive integer. 
When constructing the networks, in general, pairs of 
free sites that have $k-1$ free sites between them are connected 
iteratively \cite{juhasz}.
In the following, 
the steps of this procedure will be described in detail.

All the networks that we study are defined 
by means of aperiodic sequences, therefore, we start with a brief
introduction of them.  
The aperiodic sequences that we need are generated by substitution rules on the
letters of a finite alphabet $\mathcal{A}=\{ a,b,c,\dots\}$ that assigns a word (a finite
string of letters) $w_{\alpha}$ to each letter $\alpha\in\mathcal{A}$. 
A (finite) aperiodic sequence is
obtained by applying the inflation rule iteratively (finitely many
times) starting with a single letter (which is letter $b$ by convention).  
We use the two-letter sequence defined by the inflation rule  
\be
\sigma_{n}: \left\{
\begin{array}{c}
{ a \to w_a=aba(ba)^{n-1} }\\
{ b \to w_b=a(ba)^{n-1}, }  
\end{array}
\right. 
\label{2letter} 
\ee
where $n$ is a positive integer 
and a three-letter sequence called tripling sequence generated by 
\be
\sigma_{t}: \left\{
\begin{array}{c}
{ a \to w_a=aba} \\
{ b \to w_b=cbc} \\
{ ~c \to w_c=abc.}   
\end{array}
\right. 
\label{tripling} 
\ee
The two-letter inflation rule with $n=1$ generates the silver-mean sequence
with the first few iterations $b$,$a$,$aba$,$abaaaba$,$abaaabaabaabaaaba$,
etc, whereas with the choice $n=2$, the well-known Fibonacci sequence is generated. 

Having defined the aperiodic sequences, we return to the
construction of networks. 
First, the class of networks is defined where $k=1$, that means, neighboring
free sites are connected recursively. 
Here, a finite aperiodic sequence is chosen and the edges of 
the (finite) initial
one-dimensional lattice are labeled consecutively by the letters of this sequence.  
Assume that all the edges which are labeled by letter 
$\alpha\in\mathcal{A}$ have
a common length, denoted by $l_{\alpha}$. 
Furthermore, let the edge lengths $l_{\alpha}$ be ordered for the case of two-letter sequences as $l_b<l_a$ while
for the tripling sequence as $l_b<l_c<l_a$. 
Now, let us find the closest pair of free sites (or, to be precise,
one of the pairs with the shortest spacing between them) and connect them 
with an additional edge. 
This step is then iterated until all sites become of degree
$3$. It is clear that initially there is a multitude of pairs which
are separated by the shortest distance ($l_b$) and, in the first few steps,
these pairs are connected subsequently. Obviously, as the procedure goes on, the
minimal spacing increases and longer and longer edges will form.   
 
\begin{figure}[h]
\includegraphics[width=1\linewidth]{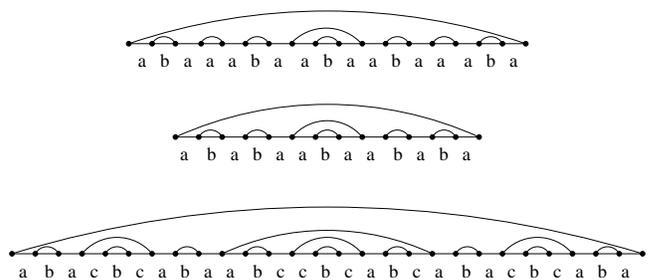}
\caption{\label{k1} Finite silver-mean, Fibonacci and tripling networks (from top to bottom).}
\end{figure}

We have studied the contact process on three different networks with $k=1$: 
the silver-mean network, the Fibonacci network and the tripling network,
which are constructed by using the corresponding aperiodic sequence. 
The structure of these networks is illustrated in Fig. \ref{k1}. 

In addition to this, we have considered networks with $k=2$, as well.
The $k=2$ tripling network and the $k=2$ silver-mean network are
constructed as follows \cite{juhasz}. 
First, the sites of a one-dimensional
lattice are labeled with the letters of the corresponding sequence. 
The sites are grouped into blocks corresponding to words $w_{\alpha}$ in the
inflation rule. 
Then, sites belonging to one-letter blocks are renamed
according to the reversed inflation rule $w_{\alpha}\to\alpha$, 
where $w_{\alpha}$ is the one-letter word corresponding to the block.      
In blocks composed of three sites, the two lateral sites are
connected, and the middle one is renamed again according to the
reversed inflation rule $w_{\alpha}\to\alpha$, where $w_{\alpha}$ is
the word corresponding to the block.     
The above step is then iterated until only one free site 
(the central one) is left.   
\begin{figure}[h]
\includegraphics[width=1\linewidth]{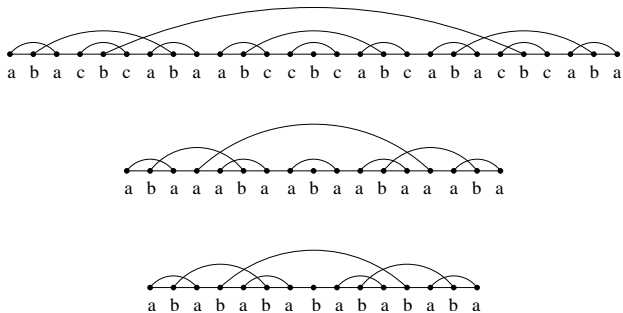}
\caption{\label{k2} Finite $k=2$ tripling, silver-mean and Hanoi-tower networks
  (from top to bottom).}
\end{figure}
The third network with $k=2$ that will be investigated is the
cubic Hanoi-tower network \cite{boettcher} that can be constructed following
the above procedure with the inflation rule  
\be
\sigma_{H}: \left\{
\begin{array}{c}
{ a \to w_a=aba }\\
{ b \to w_b=b. }  
\end{array}
\right. 
\label{Hanoi} 
\ee
The three networks with $k=2$ are illustrated in Fig. \ref{k2}.

\subsection{Diameter and random-walk dimension} 

In the rest of this section, we shall survey some intrinsic properties of the
above networks that are exactly known and are relevant with respect to the
off-critical dynamical behavior of the contact process. 
 
Beside the distance $x$ measured on the underlying one-dimensional lattice,
another metric is the shortest-path length $\ell$ between two sites which is
the minimal number of links that have to be traversed when going
from one site to the other one.  
The average length of the shortest path between two sites separated by
the distance $x$ scales in these networks as $\overline{\ell}(t)\sim x^{d_{\rm min}}$      
where $d_{\rm min}$ is the shortest-path dimension of the network. 
The diameter $D(L)$ of a typical finite graph with $L$ sites, which is the
largest shortest-path length between any two sites grows also
algebraically as $D(L)\sim L^{d_{\rm min}}$.  
The average number of sites $V(\ell)$ that can be reached in at most $\ell$
steps starting from a given site scales as $V(\ell)\sim \ell^{d_g}$
where $d_g$ is the graph dimension that is related to $d_{\rm min}$ as 
$d_{g}=1/d_{\rm min}$.

The other property that we need is the random-walk dimension of the network. 
Let us consider a continuous time random walk on (infinite) 
networks, where the walker can jump 
with unit rate to any of the sites connected with the site it resides.
The random-walk dimension $d_w$ is defined through 
the asymptotical relation 
$[\langle x^2(t)\rangle ]_{\rm typ}\sim t^{2/d_w}$,
where $x(t)$ denotes the displacement of the walker at time $t$ and  
$[\langle x^2(t)\rangle ]_{\rm typ}\equiv\exp{\overline{\ln\langle x^2(t)\rangle}}$ is the ``typical value'' of $\langle x^2(t)\rangle$.   
Here, $\langle\cdot\rangle$ denotes the average over different
stochastic histories for a fixed starting position, 
while the over-bar stands for the average over starting positions.  
Note that the expected value $\overline{\langle x^2(t)\rangle}$   
does not exist if $t>0$ since the expected value of edge lengths
is infinite (in infinite networks). This accounts for that the average of 
$\ln\langle x^2(t)\rangle$ is considered instead. 
Notice that $d_{\rm rw}$ is a dynamical exponent
that relates time and length scale of random walks.

The dimensions $d_{\rm min}$ and $d_{\rm rw}$ of 
the networks defined in the previous section have been exactly 
calculated \cite{boettcher,juhasz}; the
corresponding numerical values are shown in Table~\ref{table1} 
and Table~\ref{table2}.

\section{Contact process on networks}
\label{seccp}
\subsection{Definition of the model}

\begin{figure}[h]
\includegraphics[width=0.9\linewidth]{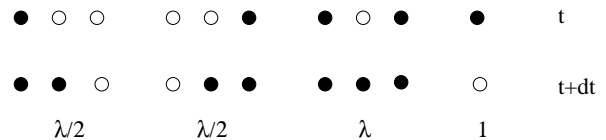}
\caption{\label{cp} Allowed transitions and the corresponding rates 
in the one-dimensional contact process. Full(empty) circles symbolize
active(inactive) sites.}
\end{figure}

The contact process is one of the earliest and simplest lattice
models that belongs to the DP universality class. 
It is a continuous time Markov process on a
state space $\{0,1\}^S$, where $S$ is a finite 
or countable graph, usually $\mathbb{Z}^d$. 
A site with state $0$($1$) is called
inactive(active) or, in the context of epidemics, healthy(infected). 
The dynamics is defined by nearest-neighbor transitions that occur
independently with given rates.
In $d$ dimension, an infected site can spontaneously become healthy ($1\to 0$) 
with rate $1$ or can infect one of its neighbors ($0\to 1$) 
with rate $\lambda/(2d)$ (see Fig.~\ref{cp}).
This process is defined on the cubic networks under consideration in the way
that infected sites are healed with rate $1$ as before, whereas each
nearest-neighbor site is infected with rate $\lambda /3$, such that (for a
site of degree $3$) the total infection rate is $\lambda$.   
In numerical simulations, this process is realized
by random sequential updates: a randomly chosen infected site 
either becomes healthy with probability $1/(1+\lambda )$ 
or a neighboring site is 
attempted to be infected with probability $\lambda /[3(1+\lambda )]$.
For sites of degree $2$ (the central site in $k=2$ networks and the surface
sites in open $k=2$ networks) and for sites of degree $1$
(the surface sites in $k=1$ networks), no update is attempted with probability 
$\lambda /[3(1+\lambda )]$ and $2\lambda /[3(1+\lambda )]$, respectively.
 
Throughout the paper we measure
the time in units of Monte Carlo (MC) steps that corresponds to 
$N$ update attempts, where $N$ is the actual number of particles at the
beginning of the MC step.  

\subsection{Studied quantities} \label{studiedq}

The quantities, the time-dependence of which we have measured at
criticality, are the 
average number of active sites $N(t)$ and the survival probability
$P(t)$, which is the probability that there is at least one active
site at time $t$. Furthermore, in the case of seed simulations, 
when initially a single site denoted by $m$ is active and all other sites are inactive, we have also measured
the second moment of the distance of the growing cluster with respect
to the origin $m$, the so called spread $R_m^2(t)$. To be precise, this
quantity  is defined as $R_m^2(t)=\langle
\sum_{i=1}^Ln_i(t)x_i^2(t)\rangle/N(t)$, where $x_i$ is the distance between
site $i$ and the origin $m$, whereas $n_i$ is a binary variable which 
is one(zero) for active(inactive) sites.  
In the case of a single initial seed, the above quantities are expected to 
follow power-laws asymptotically: 
\beqn
N_m(t)\sim t^{\eta_m} \label{Nt} \\
P_m(t)\sim t^{-\delta_m} \\ 
R_m^2(t)\sim t^{2/z_m}. 
\eeqn
In an inhomogeneous system the critical exponents $\eta_m$, $\delta_m$ and $z_m$ may be different for different initial active sites $m$. We shall see, that 
this is indeed the case, at least for the former two exponents. 
We have probed three different positions for the initial seed. 
First, it has been located at the central site(s), i.e. 
site $(L+1)/2$ for $k=2$
networks, where the number of sites $L$ is odd, and sites $L/2$ and $L/2+1$
for $k=1$, where $L$ is even. 
We shall refer to this arrangement of the initial seed by the index '$0$'. 
Second, the initial seed has also been located at the site from which the
longest edge of the (finite) network emanates. (Note that the edge connecting
site $1$ and site $L$ in Fig. \ref{k1} is not the longest one since 
they are neighbors on the ring.) 
This initialization is indexed by '$l$'. Note that in this case, as the process
starts, the spread jumps immediately to $O(x^2)$, where $x$ is the length
of the longest edge. To ensure the smooth increase of the spread, we have
modified its definition so that $x_i$ is the minimum of the distances measured 
from the two sites which are connected by the longest edge.
Third, we have considered networks built on open one-dimensional chains, where
the end sites are of degree $1$ for $k=1$ and of degree $2$ for $k=2$. 
In these networks, the initial seed has been located at the surface, i.e. 
at site $1$. The exponents corresponding to this arrangement are indexed by 's'.

In addition to this, we have measured $\overline{N(t)}$ and $\overline{P(t)}$,
the number of active sites and the survival probability, respectively, 
that are averaged over seed
simulations started from all possible initial sites $1,2,\dots,L$. The
corresponding exponents are denoted by $\eta_{\rm av}$ and $\delta_{\rm av}$. 
Note that the average of $R^2(t)$ over all possible initial sites diverges in
an infinite network for any $t>0$ since the expected value of edge-lengths is
infinite \cite{juhasz}. Nevertheless, the averaging would not provide any
new information anyway on the spread since, according to the numerical results,
the dynamical exponent $z_m\equiv z$ is independent of the location 
of the initial seed.    

Another dynamical scaling exponent characterizes the critical system 
that is started from a homogeneous, fully occupied initial state. 
In this case, the density $\rho(t)$ of infected sites 
decays asymptotically as
\begin{equation}
\rho(t) \sim t^{-\alpha}.
\label{alphadef}
\end{equation}
In case of models of the DP class defined 
on regular lattices, $\alpha=\delta$ holds due to the rapidity reversal
symmetry (see \cite{odor}). We shall, however, see that this
equality does not hold in general for the networks under study.

\section{Results}
\label{secresults}
\subsection{Off-critical behavior}

According to our numerical results, below the critical value of the
activation rate $\lambda_c$, that is characteristic for the underlying
network, the system is in the inactive phase. 
Here, the number of active sites is found to decrease exponentially in
time similar to regular lattices. The dynamics in this phase can be
essentially described by random walks of the infection since active sites
typically become rapidly inactive after activating a neighboring site.
Accordingly, the spread is well approximated by the
mean-square displacement of random walks, $R^2(t)\sim t^{2/d_{\rm rw}}$, 
with the only difference compared to regular lattices is
that the dimension $d_{\rm rw}$ entering the above relation is the
anomalous random-walk dimension of the underlying network.

In the active phase, $\lambda>\lambda_c$, the inactivation processes
are irrelevant and the infection is spreading with a constant speed
across the network. Since in $t$ time steps all sites within the distance
$\ell\sim t$ are activated, the number of active sites grows in time
as $N(t)\sim V(t)\sim t^{1/d_{\rm min}}$. As the growing cluster of
active sites is compact in this phase, the spread increases in time as 
$R^2(t)\sim t^{2/d_{\rm min}}$.

The above laws for $R^2(t)$ below and above the critical point provide
the bounds $d_{\rm min}\le z \le d_{\rm rw}$ 
for the non-trivial critical dynamical exponent $z$. Furthermore, 
the number of active sites in surviving samples must not grow faster
at criticality than in the active phase, which yields the 
inequality $\eta+\delta\le 1/d_{\rm min}$.
We shall see, that the measured critical exponents are compatible with 
these (rather weak) bounds. 

\begin{table}[h]
\begin{center}
\begin{tabular}{|l|l|l|l|l|}
\hline                  & 1D       & silver-mean & Fibonacci & tripling  \\
\hline $d_{\rm min}$    & 1        & 0.7864...   & 0.7610...& 0.6309...  \\
\hline $d_{\rm rw}$     & 2        & 1.7864...   & 1.7610...& 1.6309...  \\
\hline $L        $      &          &  665858     &  1346270  & 1594324  \\
\hline $\ln(b)$         &          & 1.7627...   &  1.4436...& 1.0986...  \\
\hline $\ln(\tau)$      &          &  2.5(1)     & 2.0(1)    & 1.4(1)     \\
\hline $\frac{\ln(\tau)}{\ln(b)}$ &&  1.4(1)     & 1.4(1)    & 1.3(1)     \\
\hline $\lambda_c$      & 3.29785  & 3.0831(4)   & 3.0146(2) & 2.79926(5) \\
\hline $\eta_0$         & 0.31368  & 0.261(3)    & 0.253(1)  & 0.192(4)   \\
\hline $\eta_l$         &          & 0.386(2)    & 0.385(2)  & 0.386(2)   \\
\hline $\eta_{\rm av}$  &          & 0.310(3)    & 0.307(3)  & 0.302(3)   \\
\hline $\eta_s$         &0.04998(2)& -0.080(2)   & -0.067(3) & -0.016(2)  \\
\hline $\delta_0$       & 0.15947  & 0.249(3)    & 0.264(3)  & 0.367(5)   \\
\hline $\delta_l$       &          & 0.124(2)    & 0.131(2)  & 0.172(2)   \\
\hline $\delta_{\rm av}$&          & 0.201(3)    & 0.207(3)  & 0.253(3)   \\
\hline $\delta_s$       &0.42317(2)& 0.590(2)    & 0.583(2)  & 0.570(2)   \\
\hline $2/z_0$          & 1.26523  & 1.424(4)    & 1.445(3)  & 1.618(6)   \\
\hline $2/z_l$          &          & 1.429(3)    & 1.453(3)  & 1.625(3)   \\
\hline $2/z_s$          &1.26523   & 1.426(5)    & 1.451(5)  & 1.624(4)   \\
\hline $\alpha$         & 0.15947  & 0.201(1)    & 0.205(1)  & 0.250(1)   \\
\hline
\end{tabular}
\end{center}
\caption{\label{table1} Various properties and estimated critical exponents of 
$k=1$ networks. The known 1D exponents are from Refs \cite{odor} and \cite{FHL01}.}
\end{table}
\begin{table}[h]
\begin{center}
\begin{tabular}{|l|l|l|l|}
\hline                  & k=2 tripling & k=2 silver-mean & Hanoi   \\
\hline $d_{\rm min}$    & 0.6309...   & 0.6232... & 0.5        \\
\hline $d_{\rm rw}$     & 1.4650...   & 1.4575... & 1.3057...   \\
\hline $L$              & 1594323     & 1607521   &  4194303    \\
\hline $\ln(b)$         & 1.0986...   & 1.7627... & 0.6931...  \\
\hline $\ln(\tau)$      & 1.4(1)     & 2.1(1)   & -   \\
\hline $\frac{\ln(\tau)}{\ln(b)}$ & 1.3(1) & 1.2(1) & -  \\ 
\hline $\lambda_c$      & 2.31269(3)  & 2.28979(3)& 2.18432(4) \\
\hline $\eta_0$         & 0.117(2)    & 0.114(4)  & 0.12(1)   \\
\hline $\eta_l$         & 0.303(5)    & 0.308(2)  & 0.33(1)    \\
\hline $\eta_{\rm av}$  & 0.296(1)    & 0.294(2)  & 0.28(1)   \\
\hline $\eta_s$         & 0.119(2)    & 0.117(2)  & 0.07(3)    \\
\hline $\delta_0$       & 0.448(4)    & 0.453(3)  & 0.49(1)   \\
\hline $\delta_l$       & 0.261(2)    & 0.259(2)  & 0.28(1)    \\
\hline $\delta_{\rm av}$& 0.268(2)    & 0.271(2)  & 0.33(1)   \\
\hline $\delta_s$       & 0.446(2)    & 0.448(2)  & 0.54(1)   \\
\hline $2/z_0$          & 1.665(2)    & 1.673(2)  & 1.90(1)   \\
\hline $2/z_l$          & 1.665(2)    & 1.675(3)  & 1.90(1)   \\
\hline $2/z_s$          & 1.667(4)    & 1.674(2)  & 1.89(1)   \\
\hline $\alpha$         & 0.269(1)    & 0.271(1)  & 0.33(1)   \\
\hline
\end{tabular}
\end{center}
\caption{\label{table2} Various properties and estimated critical exponents of
  $k=2$ networks.}
\end{table}

\subsection{Critical behavior}

The numerical simulations have been performed in networks that are built on a
finite periodic one-dimensional lattice, 
except for seed simulations started from the surface
site, where networks built on open chains have been used. 

The size (i.e. the number of sites) $L$ of aperiodic networks is not
arbitrary but it is given by the possible lengths of finite strings of the 
corresponding aperiodic sequence \cite{juhasz}. 
The system sizes that we have typically used in the simulations are shown in
Table I and II.     
The simulation time was typically $2^{18} - 2^{22}$ MC steps and 
the averaging has been performed over $10^6$ independent runs. 
In case of seed simulations, the applied system sizes were
large enough compared to the size of growing clusters 
such that the system can be regarded as infinite. 

\begin{figure}[h]
\includegraphics[width=0.9\linewidth]{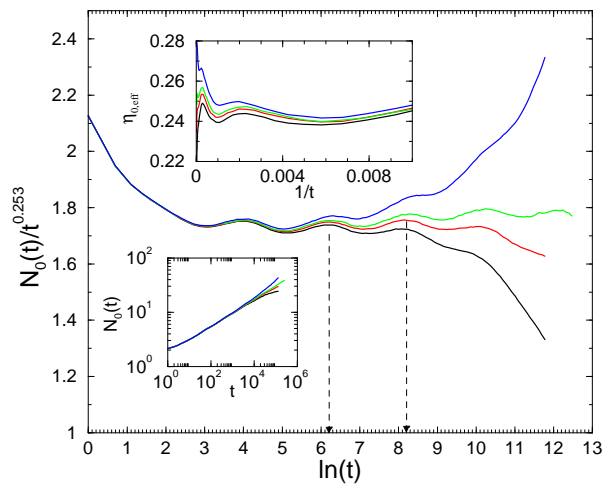}
\caption{\label{k1fe} Time-dependence of the particle number $N_0(t)$ in
 the $k=1$ Fibonacci network for
  $\lambda=3.013, 3.014, 3.0145, 3.016$ (from bottom to top).
Log-periodic modulations at criticality with a period $\ln\tau=2.0(1)$
can be read-off. The upper inset shows the corresponding local slopes;
the lower inset shows the unscaled data.}
\end{figure}

In order to estimate the critical infection rate $\lambda_c$  and 
to keep track corrections to 
scaling more clearly we have monitored the effective exponent 
$\alpha_{\rm eff}(t)$ defined by 
the local slope
\begin{equation}
\alpha_{\rm eff}(t)=-\frac{d\ln(\rho(t))}{d\ln(t)} \ .
\end{equation}
This kind of analysis helped to estimate the
other dynamical exponents ($\delta$,$\eta$,$z$), as well.
However, the presence of log-periodic oscillations made the determination 
of the critical point rather difficult, 
since they distort the monotonicity of the functions. Without these
modulations the effective exponents show upward(downward) curvature
above(below) the transition point, respectively (see Fig.~\ref{k1fe}).
In what follows, we shall illustrate the critical behavior of the
model mainly for
the particular case of the $k=1$ Fibonacci network; 
the behavior of the process on
the other networks is qualitatively similar and the corresponding quantitative
data can be found in Table I and II.

First we located the critical point by measuring
dynamical quantities and calculating the effective exponents.
In case of the $k=1$ Fibonacci network, the average number of particles 
originating from the central site increases algebraically, 
superimposed with log-periodic oscillations as can be seen in Fig.~\ref{k1fe}.
The critical point estimated by the local slopes is at $\lambda=3.0146(2)$
with moderate corrections to scaling.
The survival probability $P(t)$ and the $R^2(t)$ also show these periodic
modulations with the same period $\ln\tau=2.0(1)$. 
This is in agreement
with the expectations for critical systems with discrete spatial scale
invariance. 
Such systems remain self-similar when lengths are rescaled by a given
(non-arbitrary) scale factor $b$ that is characteristic for the system. 
The time-dependent quantities in such models are expected to 
display log-periodic oscillations with a temporal period 
 $\ln\tau$ that is related to the spatial scale factor through 
\be
\ln\tau=z\ln b,
\label{period}
\ee
 where $z$ is the dynamical exponent of the model \cite{bab}. 
The numerical values of the spatial scale factors $b$ of the networks under
study \cite{juhasz,boettcher} and the estimated temporal periods are given in
Table I and II. The data are in satisfactory agreement with Eq. (\ref{period}).   

As aforementioned, we performed simulations with three different positions of
the initial active seed. 
According to the numerical results (see Table I and II), 
in the three cases the exponents $\eta_m$ and $\delta_m$ are different, 
whereas $z_m$ and the sum $\eta_m+\delta_m$ are independent of the 
initial position $m$ \footnote{Note that this is also true for the surface critical
exponents at the ordinary transition of the one-dimensional DP \cite{FHL01}.}. 
The latter means that the growth rate of the number of active sites averaged in
surviving samples, i.e. $N_m(t)/P_m(t)$, is not influenced by the location $m$
of the initial seed. 

As opposed to this, the survival probability is sensitive to 
this circumstance.
We have found that $\delta_0>\delta_l$, which is intuitively obvious and can
be explained as follows. 
The survival probability is greatly influenced by the relative position of
the growing cluster with respect to the long edges as the active region 
reaches them. 
If the process starts from the central site, the cluster grows
typically symmetrically around the central site. Since the longer edges are 
also located symmetrically around the central site, 
the growing cluster overlaps 
with itself as in a finite system with periodic boundary conditions. 
This, however, decreases the survival probability because the rate of 
unsuccessful activation attempts is higher in an overlapping front.
Thus in this case, the long edges of the network are not utilized in a 
favorable way from the point of view of the survival of
the process.
Apparently, for initial sites which have an environment identical 
to that of the central one but only within a finite radius, 
the growth is described by the same
exponents until the cluster is within this radius.   

Contrary to this, when the process is started at a site with a long edge, 
the infection is transferred immediately to a far away place 
and the two clusters spreading out from the two sites 
connected by the long edge do not hinder each other.
Of course, in case of a long edge of length $l$, the two advancing 
fronts meet after time $t\sim O(l^z)$ and the exponents $\eta_l$ and $\delta_l$ 
describe the spreading dynamics only below this time scale. 
At this time scale, a crossover occurs to a region with a faster decaying 
survival probability. 
In fact, for a typical initial seed location, the dynamical quantities 
suffer a series of crossovers, depending on the relative position of
the initial site with respect to the longer and longer edges the cluster hits.

On the grounds of the above argumentations, 
the most favorable initial site for the cluster survival is the site
at the longest edge, while, disregarding the surface site, the most
unfavorable site is the central one. Therefore, the decay of the average
survival probability must be bounded by the surviving
probabilities in the above two extremal situations and we expect  
$\delta_0\ge\delta_{av}\ge\delta_l$ to hold.
According to our numerical results, these relations are valid indeed. 

\begin{figure}[h]
\includegraphics[width=0.9\linewidth]{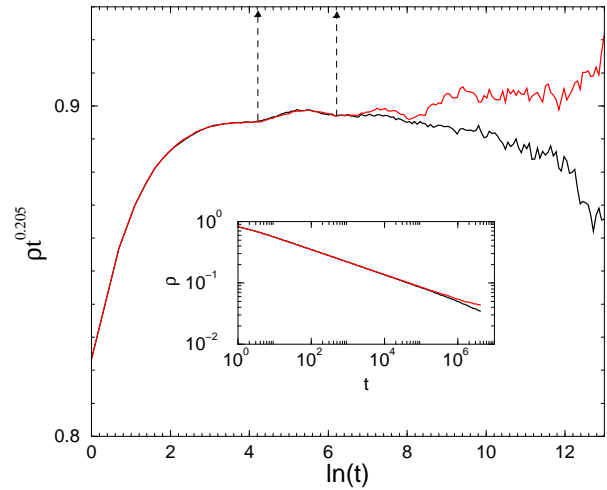}
\caption{\label{k1fa} Density decay in the $k=1$ Fibonacci network
for $\lambda= 3.0145, 3.016$ (from bottom to top).
Log-periodic oscillations can be seen as for $N_0(t)$.
The inset shows the unscaled data.}
\end{figure}

\begin{figure}[h]
\includegraphics[width=0.85\linewidth]{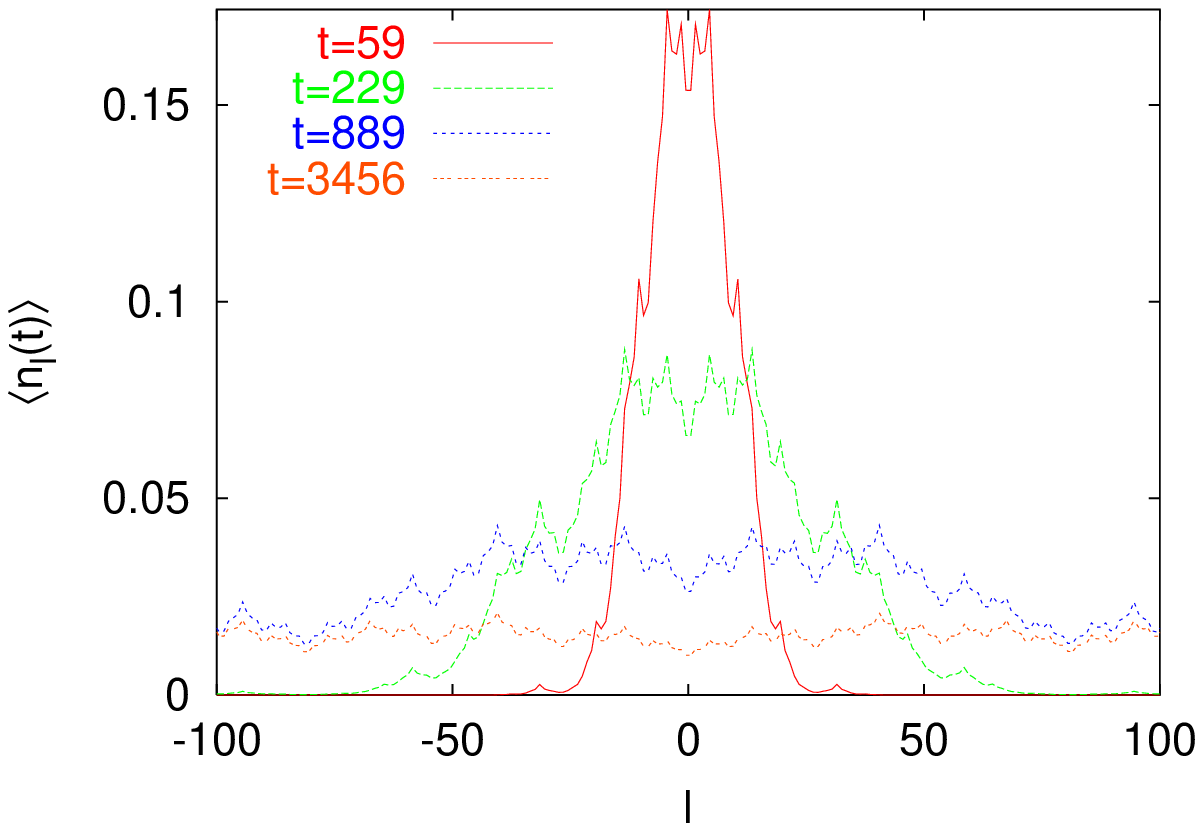}
\includegraphics[width=0.85\linewidth]{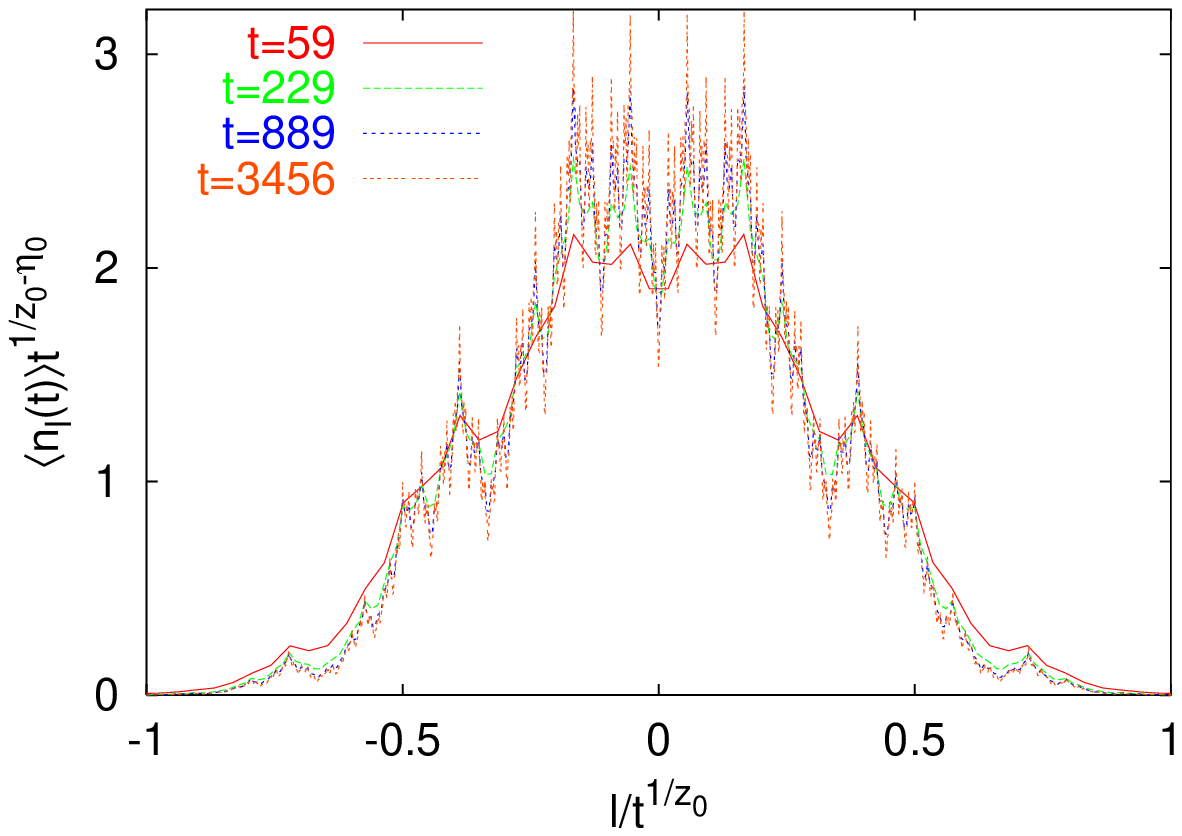}
\caption{\label{prof} Upper figure: 
Average occupation numbers $\langle n_l(t)\rangle$ 
of sites at different times, plotted against the distance $l$ measured from
the center. The logarithms of times are integer multiples of the estimated
temporal period $\ln\tau$. The process was started from
the central sites of the $k=1$ tripling network and the averaging has been
performed over $10^7$ runs. Lower figure: Scaling plot
with the same data.}
\end{figure}

Starting from a fully occupied initial state, the decay of the density at the
critical point is characterized by the exponent $\alpha$ as shown
in Fig.\ref{k1fa}.
As can be seen from the estimated exponents given in the tables, in
general, $\delta_m\neq \alpha$, in contrast with the CP on regular
lattices, where $\delta =\alpha$ holds.
As a consequence, the hyper-scaling law of DP does not hold on the
networks under study, i.e. $2\delta_m + \eta_m \neq d/z$ with $d=1$. 
But the set of critical exponents are compatible with the generalized 
hyper-scaling-law \cite{TTP}:
\be\label{genhyper}
\delta_m + \alpha + \eta_m = d/z \ .
\ee
with $d=1$. 
We thus conclude that the rapidity reversal symmetry, which
is necessary to the validity of the hyper-scaling law is broken.
In other words, $P(t)$ and $\rho(t)$ scale in a different way, which
is a consequence of the presence of long edges that render the system
inhomogeneous in space.
The inhomogeneity of the system is illustrated in Fig.~\ref{prof},
where the local particle densities $\langle n_l(t)\rangle$ are 
plotted at different times against the distance $l$ measured from the center. 
As can be seen, the regularly arranged long edges induce modulations in the profiles.
As the number of infected sites grows at criticality as given in
Eq. \ref{Nt} and the active sites are concentrated in a region 
of size $\xi(t)\sim t^{1/z}$, we expect the density profiles to have 
the following scaling form:
\be 
\langle n_l(t)\rangle=t^{\eta_m-1/z}\tilde\rho(l/t^{1/z}).
\ee
This has been plotted in Fig.~\ref{prof}. As can be seen, the number
of peaks in the profiles is increasing with $t$ and, as a consequence,
the scaling function $\tilde\rho(x)$ is non-smooth.    

Nevertheless, the numerical results suggest that $\delta_{av}=\alpha$
is valid within 
error margin, thus for the average cluster-spreading exponents the 
hyper-scaling of DP is fulfilled, i.e.
\be
2\delta_{av} + \eta_{av} = 1/z.
\ee
This indicates that the breaking of rapidity reversal symmetry  
is indeed related to the presence of the spatial inhomogeneities in the 
system. This is in agreement with field theory of directed percolation 
with long-range spreading \cite{JS08}, where the rapidity reversal symmetry persists.

\begin{figure}[h]
\includegraphics[width=0.9\linewidth]{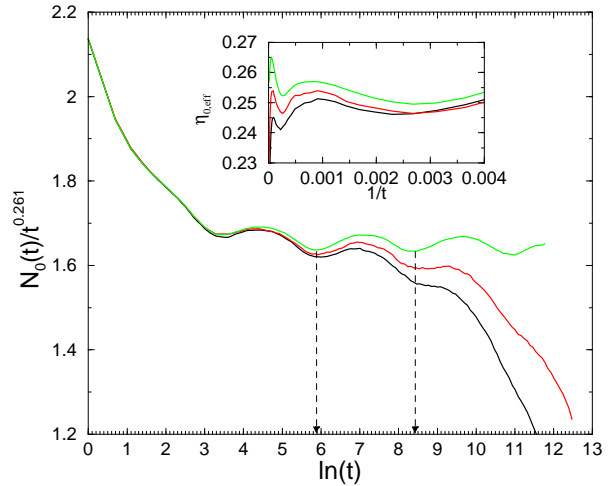}
\caption{\label{k1se} Time-dependence of the particle number $N_0(t)$ 
in the $k=1$ silver-mean network for $\lambda= 3.081, 3.082, 3.083$ 
(from bottom to top). The period of oscillations is $\ln\tau= 2.5(1)$.
The inset shows the effective exponents.}
\end{figure}

\begin{figure}[h]
\includegraphics[width=0.9\linewidth]{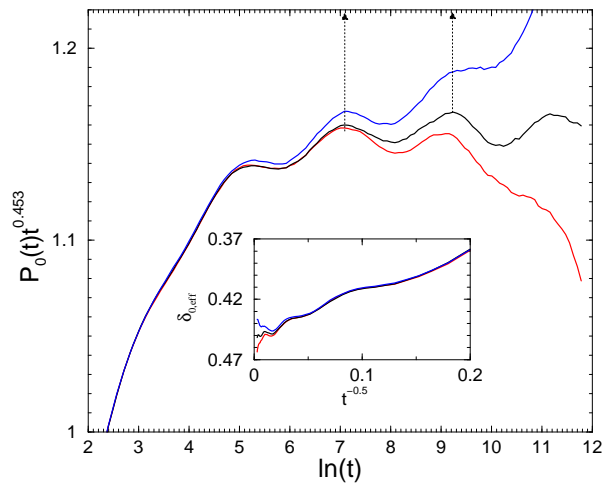}
\caption{\label{k2sd} Time-dependence of the survival probability $P_0(t)$ in the $k=2$ silver-mean network for $\lambda =2.2896,2.28979,2.29$ 
(from bottom to top). The period of oscillations is $\ln\tau= 2.1(1)$.
The inset shows the effective exponents.}
\end{figure}

\begin{figure}[h]
\includegraphics[width=0.9\linewidth]{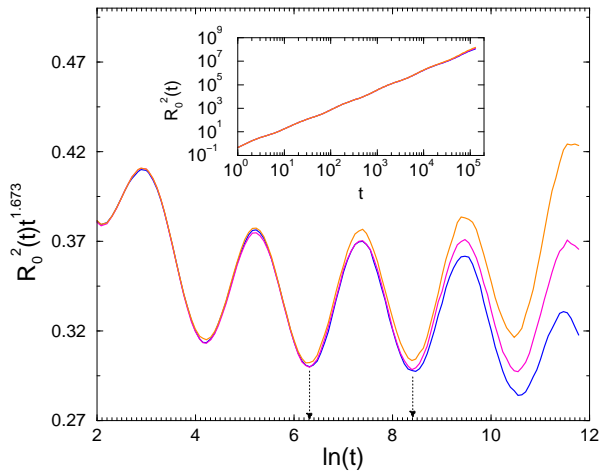}
\caption{\label{k2sz} Time-dependence of the spread $R^2_0(t)$ in the $k=2$ silver-mean network for $\lambda =2.2896,2.28979,2.29$ (from bottom to top).
The inset shows the unscaled data.}
\end{figure}

Similar to the $k=1$ Fibonacci network, 
we performed the above analysis for the $k=1,2$
silver-mean and tripling networks and for the $k=2$ Hanoi-tower network
(see Figs. \ref{k1se},\ref{k2sd} and \ref{k2sz}). 
The contact process on these networks exhibits the same qualitative
features as on the $k=1$ Fibonacci network except that, for the Hanoi-tower
network, log-periodic oscillations cannot be observed presumably owing to their
small amplitude. The estimates of critical exponents, 
which depend on the underlying network, can be found in Table I and II.

\section{Discussion}
\label{secdiscussion}

First, we give a brief summary of the results obtained so far.
In each network, a phase transition between active and inactive phases can be
identified at some finite value of the control parameter by inspecting 
dynamical properties such as the time dependence of the number of active
sites, the survival probability and the spread.
At the absorbing phase transition, conventional power-law dependence 
of the above quantities can be observed apart from log-periodic oscillations
that are related to the discrete scale invariance of the underlying networks.
In a given network, the critical exponents $\eta_m$ and $\delta_m$ 
depend on the location of the initial seed. 
The cluster exponents satisfy the generalized hyper-scaling relation. 
However, in case of averaging over runs started at different initial 
seed coordinates, the hyper-scaling relation of DP holds, too. 
This means that the rapidity reversal symmetry is broken as a
consequence of the spatial inhomogeneity.
The dynamical critical exponents of the phase transition differ from
that of the one-dimensional DP universality class and are found to be
characteristic for the underlying network.   

\begin{figure}[h]
\includegraphics[width=0.9\linewidth]{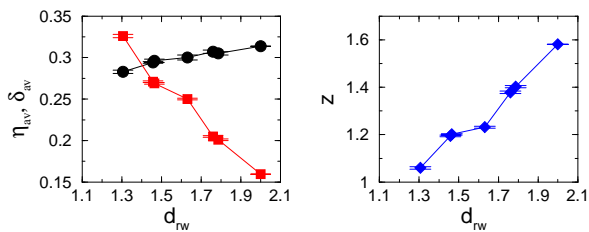}
\caption{\label{exps} The critical exponents $\eta_{av}$ ({\Large $\bullet$}), 
$\delta_{av}=\alpha$ ($\blacksquare$) and $z$ ($\blacklozenge$) of the six studied
systems and the one-dimensional CP plotted against $d_{\rm rw}$ of 
the underlying network.}
\end{figure}

We close this work by discussing the relation between the critical
exponents of the CP and the shortest-path dimension 
(or random-walk dimension) of the underlying network.
The measured critical exponents of the six network models are plotted
against the random-walk dimension of the corresponding network 
in Fig. \ref{exps}. 
For decreasing $d_{\rm rw}$ the critical exponents $\eta_{av}$ and
$\delta_{av}=\alpha$ move towards the mean-field values of DP 
($\eta_{MF}=0$, $\delta_{MF}=1$) almost monotonically.
These exponents lie in between the corresponding values of the 
one-dimensional and the two-dimensional DP universality classes.
Contrary to this, the dynamical exponent $z$ does not move toward the
mean field value $z_{MF}=2$, but decreases with decreasing $d_{\rm rw}$. 
Thus, it moves parallel with the dynamical exponent of the random
walk.

We recall that the $k=1$ silver mean and $k=1$ Fibonacci network are
the first two members of a family of networks, which are defined by
the inflation rule in Eq.~(\ref{2letter}), and which are
parameterized by an integer $n$.
For these networks, it is known that $d_{\rm min}\to 1$ and 
$d_{\rm rw}\to 2$ when $n\to\infty$, i.e. in this limit, the
characteristics of the regular one-dimensional lattice are 
recovered \cite{juhasz}. Therefore, we conjecture that the critical
exponents of CP defined on these networks approach the one-dimensional
DP values without limits as $n\to\infty$. 
In the opposite limit, i.e. when $d_{\rm rw}\to 1$ 
(and $d_{\rm min}\to 0$), it is an open question what are the 
limiting values of the critical exponents of the CP. 

Finally, we mention that the above observations open up the possibility 
to design networks on which dynamical processes evolve 
in a prescribed way.
The optimization of spreading or transport processes in
networks is of great practical relevance \cite{kleinberg,mcculloh,PSV}. 
Besides the tuning of degree-distribution or the dependence of
transition rates on the degrees of sites in non-regular (such as
scale-free) networks \cite{giuraniuc,kji,yang}, 
the networks studied in this work offer an alternative way of
controlling critical dynamics by means of choosing appropriate sets
of long links and the findings obtained here
may provide ideas in optimization problems.    

\acknowledgments
We thank M. Henkel and H. Park for the useful comments.
This work has been supported by the Hungarian National Research Fund
under Grant No. OTKA K75324. The authors thank for the access to HUNGRID.

\end{document}